# Effect of C additives with 0.5% in weight on structural, optical and superconducting properties of Ta–Nb–Hf–Zr–Ti high entropy alloy films

Tien Le [1, 2, a)], Yeonkyu Lee [3, a)], Dzung T. Tran [2], Woo Seok Choi [2], Won Nam Kang [2], Jinyoung Yun [3], Jeehoon Kim [3, b)], Jaegu Song [4], Yoonseok Han [4], Tuson Park [4], Duc H. Tran [5], Soon-Gil Jung [6, b)], Jungseek Hwang [2, b)]

### AFFILIATIONS

[1] University of Science and Technology of Hanoi, Vietnam Academy of Science and Technology, 18 Hoang Quoc Viet, Cau Giay, Hanoi, 100000, Viet Nam

[2] Department of Physics, Sungkyunkwan University, Suwon, 16419, South Korea

[3] Department of Physics, Pohang University of Science and Technology, Pohang, 37673, South Korea

[2] Department of Physics, Pohang University of Science and Technology, Pohang, 37673, South Korea

[4] Center for Quantum Materials and Superconductivity (CQMS), Department of Physics, Sungkyunkwan University, Suwon, 16419, South Korea

[5] Faculty of Physics, University of Science, Vietnam National University, Hanoi, 100000, Viet Nam

[6] Department of Physics Education, Sunchon National University, Suncheon, 57922, South Korea

a) These authors contributed equally to this work

b) Corresponding authors

Email: jeehoon@postech.ac.kr (Jeehoon Kim), sgjung@scnu.ac.kr (Soon-Gil Jung), jungseek@skku.edu (Jungseek Hwang)

## ABSTRACT

We investigated the superconducting (SC) properties of Ta–Nb–Hf–Zr–Ti high-entropy alloy (HEA) thin films with 0.5% weight C additives. The C additives stabilize the structural properties and enhance the SC critical properties, including $\mu_0H_{c2}$ (13.45 T) and $T_c$ (7.5 K). The reflectance of the C-added HEA film is enhanced in the low-energy region, resulting in a higher optical conductivity, which is consistent with the lower electrical resistivity. In addition, we observed SC vortices in the C-added HEA film using magnetic force microscopy. The magnetic penetration depths ($\lambda$) of the pure HEA and C-added HEA films were estimated from their Meissner force curves by comparing them with those of a reference Nb film. At 4.2 K, the $\lambda$ of the C-added film is 360 nm, shorter than that of the pure HEA film (560 nm), indicating stronger superconductivity against an applied magnetic field.

**1. Introduction**

In 2004, Yeh et al. were the first to use the term high-entropy alloys (HEAs), now the accepted name alloys with more than four or five elements in equimolar or near-equimolar ratios [1-3]. Due to the high entropy of element mixing in HEAs, simple phases such as body-centered cubic (bcc), face-centered cubic (fcc), and hexagonal close-packed (hcp) are more easily observed than other complex phases [4-7]. The development of HEAs has opened new frontiers in materials science, offering exceptional physical and mechanical properties compared to traditional alloys. These properties include superior hardness, high strength, flexibility, fracture toughness, thermal stability, enhanced corrosion resistance, and irradiation resistance [4, 6, 8-11]. Among these, Ta–Nb–Hf–Zr–Ti HEAs have emerged as fascinating candidates because, in addition to the aforementioned properties, they also display type-II superconductivity [12, 13]. This phenomenon can be attributed to BCS-type phonon-mediated superconductivity in the weak electron-phonon coupling limit. Additionally, as expected for highly disordered superconductors, HEAs exhibit a large upper critical field and reduced superfluid density [12].

Recently, Jung et al. reported that $Ta_{1/6}Nb_{2/6}Hf_{1/6}Zr_{1/6}Ti_{1/6}$ films produced by pulsed laser deposition have remarkably high critical current densities ($J_c$), exceeding 1 MA cm$^{-2}$ at 4.2 K and the highest critical temperature ($T_c$) reached 7.28 K [10]. The sintering and annealing temperatures significantly affect the values of $T_c$ due to the instability and distortion of the crystal lattices [10, 12, 14-17]. In addition, in magnetization–magnetic field ($M$–$H$) hysteresis loops, substantial flux jumps at the lower field, indicative of thermomagnetic instability, hindered the observation of the lower critical field ($\mu_0 H_{c1}$) which is related to the value of the magnetic penetration depth ($\lambda$) [10].

In the report, we investigate the effect of C additives on the crystal structure and superconducting properties of Ta–Nb–Hf–Zr–Ti HEA films. The addition of small-radius elements, such as C or N, has been recognized as an effective means of improving the strength of steel and other traditional alloys. C and N, which possess small radii of 67 pm and 65 pm, respectively, are commonly utilized as interstitial dopants in Fe–Co–Cr–Ni–Mn and Ta–Nb–Hf–Zr–Ti HEAs [18-20]. The HEA films were deposited on $c$-cut sapphire by pulsed laser deposition (PLD) at 520 °C. Interestingly, the HEA film with the 0.5% C additives showed an improvement in superconductivity. Furthermore, superconducting vortices were observed in the HEA film containing 0.5% weight C additives in a low-temperature magnetic force microscopy (MFM) image. In contrast, the pure HEA film does not exhibit clear vortices. Moreover, MFM was utilized to determine the values of $\lambda$ by directly comparing the Meissner force curves of the sample under investigation with those obtained for a reference Nb sample with a well-established $\lambda$ of 110 nm.

## 2. Experiment

The initial high-purity Ta, Nb, Hf, Zr, and Ti were precisely weighed in a predetermined compositional ratio of 1:2:1:1:1 and thoroughly mixed. Subsequently, the amalgamated mixture was subjected to arc-melting under a high-purity Ar partial pressure to prepare a $Ta_{1/6}Nb_{2/6}Hf_{1/6}Zr_{1/6}Ti_{1/6}$ HEA target with a diameter of 10 mm. The methodology employed to prepare the target followed the procedure outlined for the synthesis of bulk samples. A Ta–Nb–Hf–Zr–Ti target with 0.5% weight of C additives was prepared using the same technique with high-purity C powder.

Ta–Nb–Hf–Zr–Ti films were grown on Al$_2$O$_3$(0001) substrates by PLD from the prepared targets. The laser beam was produced utilizing a KrF excimer laser (wavelength of 248 nm, IPEX864; LightMachinery), and the deposition of thin films occurred in a PLD chamber with a base pressure of approximately $1 \times 10^{-6}$ Torr. During the growth, the substrate temperature was maintained at 520 °C, and the laser energy density was roughly 4 J.cm$^{-2}$ with a repetition rate of 10 Hz. The number of laser pulses was fixed at 20,000 for all the samples. The pure and 0.5% weight C-added HEA films are denoted as HEA and HEAC-0.5%, respectively. The film thickness is approximately 345 nm.

The crystal structures of the HEA films were examined utilizing X-ray diffraction (XRD, Miniflex-600) with Cu-Ka radiation. The surface morphologies and uniformity of the samples were investigated using scanning electron microscopy (SEM) and energy-dispersive X-ray spectroscopy (EDS). The electrical resistivity ($\rho$) under magnetic fields was measured using a standard four-probe method within a physical property measurement system (PPMS 9 T, Quantum Design). The optical spectra were measured by Fourier transform infrared (FT-IR) spectroscopy (Burker Vertex 80v) from 100 to 40,000 cm$^{-1}$. The optical conductivity spectra were extracted from the measured reflectance spectra ($R(\omega)$) using Kramers-Kronig analysis [21]. This process required extending the measured reflectance data beyond the observed range to both low and high frequencies. To extrapolate to zero frequency, we applied the Hagen-Rubens relation ($1 - R(\omega) \propto \sqrt{\omega}$). For the high-frequency extrapolation, we assumed $R(\omega) \propto \omega^{-1}$ from 40,000 to $10^6$ cm$^{-1}$ and $R(\omega) \propto \omega^{-4}$ beyond $10^6$ cm$^{-1}$, which corresponds to a free-electron response [22]. MFM measurements were conducted using a home-built $^4$He MFM probe in the temperature range from 4.25 K to 300 K [23]. All the measurements were performed using the same MFM tip (PPP-MFMR, NANOSENSORS$^{TM}$). The force gradient $\partial F/\partial z$ is acquired from the frequency shift $\Delta f$ using the equation $\frac{\partial F}{\partial z} = -2k\frac{\Delta f}{f_0}$, where $k$ (2.7 N/m) is the force constant and $f_0$ (73.264 kHz) is the resonance frequency of the MFM cantilever [24].

## 3. Results and Discussion

In the $Ta_{1/6}Nb_{2/6}Hf_{1/6}Zr_{1/6}Ti_{1/6}$ HEAs, the elements are arranged in a body-centered cubic (BCC) lattice, as depicted in Fig. 1(a), with C appearing much smaller than the other elements. Consequently, the C additives are expected to contribute to the lattice as interstitial defects. The XRD pattern of $\Theta$-$2\Theta$ scans in Fig. 1(b) shows that the positions of the (110) and (220) peaks in both HEA and HEAC-0.5% samples barely change, indicating that the lattice constants of the BCC structural samples remain at 3.377 Å and 3.379 Å, respectively. The incorporation of carbon interstitials also raises the mixing entropy, contributing to the stabilization of the single BCC phase in some HEAs [25]. Zhu et al. also reported that the slight extension of the lattice constant and volume of the medium-entropy alloy (MEA) $MoReRuC_x$ results from C interstitials with a low concentration below x=0.1 [26, 27]. Our result is similar to those in the previous studies [26, 27]. The SEM images (Fig. 2(a)-(b)) show the smooth surface morphologies of HEA and HEAC-0.5%. The HEA film exhibited a smoother surface than the HEAC-0.5%. Additionally, several droplets originating from the target during deposition were observed on the surface. The thickness of the films is approximately 345 nm, as observed in the cross-sectional SEM image (Fig. 2(c)). Using EDS, we verified that the elemental proportions in the films closely corresponded to those of the target material (Table 1). Notably, EDS can detect carbon but lacks accuracy for light elements [28, 29]. Fig. 2(d)-(e) show EDS mappings of the HEA and HEAC-0.5% samples, indicating a homogeneous distribution of all elements.

To investigate the superconductivity of the samples, the temperature-dependent resistivity ($\rho$) was measured from 2 to 10 K under various static magnetic fields, as shown in Fig. 3(a). The critical transition temperature ($T_{cR}$) was chosen to correspond to a 50% reduction in resistivity during the superconducting transition. At zero magnetic field, the $T_{cR}$ of the pure HEA film is 7.25 K, which is close to the highest $T_{cR}$ = 7.28 K reported by Jung et al. for an HEA film fabricated at 520 °C [10]. The fabrication of the target in this study utilized the arc-melting (AM) method, whereas Jung et al. utilized spark plasma sintering (SPS) [10]. The $T_c$ of bulk targets fabricated using the AM and SPS methods were 7.85 and 7.83 K, respectively, demonstrating similarity in the $T_c$ obtained from the films [30-32]. For the HEAC-0.5% sample, the $T_{cR}$ increases to 7.50 K at zero magnetic field. Typically, additional elements introduced into the original superconductors often become defects in the crystal lattice, impeding the formation of Cooper pairs and leading to a decrease in $T_c$. However, the added C atom does not replace other elements but instead acts as an interstitial atom, stabilizing the crystal lattice and thereby improving $T_c$.

From the temperature-dependent electrical resistivity under magnetic fields perpendicular to sample surfaces (0 to 9 T), we obtained the upper critical fields $\mu_0H_{c2}(T)$ using the criterion for 50% superconducting transition, as shown in Fig. 3(b). The dashed black and dot-dashed red lines represent the linear slopes for $\mu_0H_{c2}$ near $T_{cR}$, with $d(\mu_0H_{c2})/dT$ near $T_{cR}$ being -2.54 T/K and -2.59 T/K for HEA and HEAC-0.5%, respectively. The upper critical fields at 0 K, denoted as $\mu_0H_{c2}(0)$, are determined using the Werthamer–Helfand–Hohenberg (WHH) formula under the dirty limit [33, 34]: $\mu_0H_{c2}^{WHH}(0) = -0.693 T_{cR}(d(\mu_0H_{c2})/dT)_{T_{cR}}$. At the zero-temperature limit, the upper critical fields $\mu_0H_{c2}(0)$ for HEA and HEAC-0.5% are 12.73 T and 13.45 T, respectively. The enhanced $\mu_0H_{c2}(0)$ of the HEAC-0.5% sample compared to the HEA sample is attributed to the higher $T_c$ of the HEAC-0.5% sample, as the slopes of $(d(\mu_0H_{c2})/dT)_{T_{cR}}$ remain unchanged. Furthermore, the superconducting coherence lengths ($\xi$) can be estimated using $\mu_0H_{c2}(0) = \Phi_0/2\pi\xi^2(0)$, where $\Phi_0$ is the flux quantum (2.07×10$^{-15}$ Wb) [35]. The estimated coherence lengths ($\xi(0)$) of the HEA and HEAC-0.5% samples are 5.08 nm and 4.95 nm, respectively.

Fig. 4 shows the measured resistivity $\rho(T)$ in a temperature range from 2 to 300 K. The residual-resistivity ratio (RRR = $\rho(300 K)/\rho(8K)$) of HEA is 1.15, similar to that of AM and SPS bulk samples, but is lower than the RRR of 1.57 for HEAC-0.5% [30, 31]. In metals and alloys, resistivity arises from collisions between conduction electrons and oscillating lattice atoms (phonons), impurities, and imperfections. The temperature dependence of the resistivity $\rho$ in the normal state can be described by the Bloch–Grüneisen formula [36, 37]:

$$\rho(T) = \rho_0 + A(T/\Theta_{Debye})^5 \int_0^{\Theta_{Debye}/T} \frac{x^5}{(e^x-1)(1-e^{-x})} dx,$$

where $\rho_0$ is the residual resistivity, $A$ is the pre-factor, and $\Theta_{Debye}$ is the Debye temperature. By fitting temperature-dependent resistivity curves in the normal state, the $\Theta_{Debye}$ values of the HEA and HEAC-0.5% samples were estimated at approximately 267.3 and 275.6 K, respectively. The typical $\Theta_{Debye}$ value for HEA has been reported as approximately 200-300 K [30, 31, 38]. The electron-phonon coupling constants ($\lambda_{el-ph}$) of both HEA and HEAC-0.5% samples were approximately 0.75, obtained from the McMillan equation based on the BCS-Eliashberg theory given by the following formula [34, 39, 40]:

$$\lambda_{el-ph} = \frac{\mu^* \ln\left(\frac{1.45 T_{cR}}{\Theta_{Debye}}\right) + 1.04}{\ln\left(\frac{1.45 T_{cR}}{\Theta_{Debye}}\right)(1-0.62\mu^*) - 1.04},$$

where $\mu^* = 0.13$ is the Coulomb pseudopotential for intermetallic superconductors. The electron-phonon coupling constant value is valid when $\lambda_{el-ph}$ is less than the theoretical maximum value, 1.25 [34]. Superconductors with $\lambda_{el-ph}$ significantly less than 1 are classified as weak coupling; those with $\lambda_{el-ph}$ approximately equal to 1 are intermediate coupling; and those with $\lambda_{el-ph}$ greater than 1 are strong coupling [41]. The $\lambda_{el-ph}$ values of HEA and HEAC-0.5% films are within the range of intermediate coupling, yet they are smaller than those of AM and SPS bulk samples, suggesting weaker coupling in the films [30-32].

Fig. 5(a) displays measured reflectance spectra ($R(\omega)$) of HEA and HEAC-0.5% at 300 K in a wide spectral range from far infrared to ultraviolet. Both reflectance spectra show metallic behavior; as the frequency decreases, the reflectance approaches 1.0 in the low frequency region. The reflectance of HEAC-0.5% is larger than that of HEA in the whole measured range. There are two broad dips around 5000 and 15,000 cm$^{-1}$, which might be associated with charge localization and an interband transition, respectively. Fig. 5(b) and (c) show the optical conductivity spectra of HEA and HEAC-50%, respectively, obtained from the measured reflectance spectra using a Kramers-Kronig analysis [21]. Both conductivity spectra show finite values near zero frequency, confirming their metallic behavior. There is a broad dip around 13,000 cm$^{-1}$, where the interband transitions may set in. For more quantitative analysis, we fitted the optical conductivity using a Drude-Lorentz model. In the Drude-Lorentz model formalism, the real part of the optical conductivity ($\sigma_1(\omega)$) is expressed as:

$$\sigma_1(\omega) = \frac{\Omega_{p,D}^2}{4\pi} \frac{1/\tau_{imp}}{\omega^2 + [1/\tau_{imp}]^2} + \sum_i \frac{\Omega_{p,i}^2}{4\pi} \frac{\omega^2 \gamma_i}{(\omega^2 - \omega_i^2)^2 + (\omega \gamma_i)^2}$$

where $\Omega_{p,D}$ and $1/\tau_{imp}$ are the Drude plasma frequency and impurity scattering rate, respectively, while $\Omega_{p,i}$, $\omega_i$, and $\gamma_i$ are the plasma frequency, the resonance frequency, and the damping parameter of the *i*th Lorentz mode. We fitted the optical conductivity with one Drude and two Lorentz modes (Lorentz 1 and Lorentz 2), as shown in Fig. 5(b) and (c). The fitting parameters are shown in Table 2. The Drude model describes the charge carriers in a material system with two parameters: the plasma frequency and scattering rate. The square of the plasma frequency is proportional to the charge carrier density. The charge carrier density of the HEAC-0.5% sample is around 2.9 times that of the HEA sample, indicating that the C additives increase the charge carrier density. The impurity scattering rate of the HEAC-0.5% sample is 1.7 times that of the HEA sample, indicating that C additives may obstruct the flow of charge carriers. Overall contribution to the DC conductivity (or resistivity: $\rho$) can be written as $\rho \equiv \frac{1}{\sigma_0} = \frac{4\pi}{\Omega_{p,D}^2} \frac{1}{\tau_{imp}}$, where $\sigma_0$ is the DC conductivity. The resistivities of the HEAC-0.5% and HEA estimated from the Drude fitting parameters are 92 and 149 μΩcm at 300 K, respectively, which are consistent with the measured resistivities. Lorentz 2 is associated with the interband transitions, but the Drude and Lorentz 1 modes are associated with the intraband transitions. As we explained previously, the Drude mode describes the charge carriers in the samples. We propose that Lorentz 1 can be related to the localized charge carriers caused by the strong disorders in the HEA samples.

MFM measurements were conducted to observe superconducting vortices and estimate the magnetic penetration depth $\lambda$. Figs. 6(a)–6(c) present the MFM images of the Nb, HEA, and HEAC-0.5% films in the superconducting state at 4.25 K. We detected vortices in the Nb and HEAC-0.5% films, which were attributed to stray fields from the MFM instrument. These observations reveal that the spaced vortices form a disordered lattice, indicative of pinning effects. Notably, the HEA sample showed no separate vortices, underscoring the influence of C addition on the pinning force, as evidenced by the presence of well-separate vortices in the HEAC-0.5% sample (Fig. 6(c)).

A comparative analysis of the line profiles across the vortices, indicated by the white lines in Figs. 6(a) and 6(c), is depicted in Fig. 6(d). The scaled line profile of a vortex in the HEAC-0.5% film (dashed blue curve) is almost identical to the line profile of a vortex in the Nb film (black curve), suggesting a common origin for the circular features in the MFM images, namely, the vortices. The reduced absolute amplitude of the vortex line profile of the HEAC-0.5% film compared to that of the Nb film indicates a greater penetration depth $\lambda$ for HEAC-0.5%, as evidenced by a weaker repulsive Meissner force on the tip at similar lift heights.

We investigated the magnetic penetration depth ($\lambda$) using a comparative method with a well-characterized reference sample: a 300 nm Nb film with $\lambda_{Nb}$ = 110 nm at 4.2 K [42-44]. This method involved comparing the Meissner force curve at the pure Meissner state position, as indicated by the red dot in Fig. 6(a) for the reference Nb film, with those obtained at equivalent pure Meissner state positions, marked by the red dots in Figs. 6(b) and 6(c) for the studied samples. The expression for the Meissner force curve detected by the MFM tip is given by: $\frac{\partial F}{\partial z} = \frac{B\Phi_0}{(z+\lambda)^3}$, where $B$ is a pre-factor reflecting the sensor's geometry and the magnetic moment, $z$ is the distance between the tip and the sample surface, and $\lambda$ is the in-plane penetration depth [42]. To determine $\lambda$ for HEA samples, we used the formula: $\lambda = \lambda_{Nb} + \Delta z$, where $\Delta z$ is the shifted distance required to align the Meissner force curves of the HEA samples with that of the reference Nb along the $z$-axis. Applying this method, we calculated $\lambda_{HEA}$ (4.25 K) as 570 nm, derived from $\lambda_{Nb}$ (4.25 K) + $\Delta z$ = 110 nm + 460 nm, and $\lambda_{HEAC-0.5\%}$ (4.25 K) as 320 nm, derived from $\lambda_{Nb}$ (4.25 K) + $\Delta z$ = 110 nm + 210 nm. The significantly shorter $\lambda$ (320 nm) of the HEAC-0.5% film compared to that (570 nm) of the HEA film at 4.25 K indicates enhanced superconductivity in the HEAC-0.5% sample. Because the HEA sample is an $s$-wave superconductor [45], we can apply the Gorter–Casimir approximation: $\lambda(T) = \frac{\lambda(0)}{\sqrt{1-(T/T_c)^4}}$ [35, 46]. From this, the values of $\lambda_{HEA}(0)$ and $\lambda_{HEAC-0.5\%}(0)$, which are 532.48 nm and 302.09 nm, respectively. Ginzburg-Landau theory suggests that in thin films, as disorder is reduced, the penetration depth ($\lambda$) decreases and the coherence length ($\xi$) increases, indicating a shift from a dirtier to a cleaner superconducting limit. In this case, $\lambda\xi \approx$ constant, as the thermodynamic critical field ($\mu_0 H_c$) is expected to remain unaffected as $T_c$ remains unchanged. However, for the thin films studied in this work and considering $\mu_0 H_c = \frac{\Phi_0}{2\sqrt{2}\pi\xi(0)\lambda(0)}$ [35], we obtained $\mu_0 H_c$ values of 0.081 T and 0.147 T for HEA and HEAC-0.5%, respectively. This change in the HEA, where disorder is typically due to the scattering of substitutional elements at the nanometer scale, indicates unexpected behavior. As shown, the superconducting parameters are strongly affected by interstitial impurities, beyond the expectations for dirty superconductors.

The changes in the $\lambda$ values discussed above are also evident in the vortex visualization by MFM images of the different films. Figure 7 shows the results for the HEA (panels a-c) and HEAC-0.5% (panels d-f) obtained at various temperatures below their $T_{cR}$'s. Figure 7(a) at 4.25 K for the HEA reveals dark-colored droplets, as seen in the SEM images, along with a cluster of vortices, depicted in yellow. This clustering indicates inhomogeneities in the sample that cause the accumulation of vortices in specific areas. Further evidence of weak vortex pinning was obtained by manipulating vortex clusters using the magnetic moment of the tip. As the temperature increased to 5 K, the force gradient intensity of the cluster decreased, and manipulating the vortices became easier (Fig. 7(b)). At 6 K, the cluster and manipulation effects dissipated, probably due to the activated thermal vortex motion, leaving only the topographical signals of the droplets (Fig. 7(c)).

In contrast, for the HEAC-0.5% sample, vortices were present at 4.25 K, as shown in Fig. 7(d). At 5 K, vortices still existed, and manipulation of the vortices occurred (Fig. 7(e)). At 6 K, the vortices largely disappeared, and manipulation was the predominant feature (Fig. 7(f)). The differences in vortex behavior may be related to variations in the pinning landscape and intrinsic properties of the samples. Because the pinning energy $U_p$ depends on the thermodynamic critical field and $\xi$, as $U_p \sim \frac{(\mu_0 H_c)^2}{8\pi}(\frac{4}{3})\xi^3$ [47], larger vortex pinning is expected for the HEAC-0.5% sample. As the temperature increased to 5 and 6 K, the thermally activated vortex depinning led to a decrease in the force gradient and pinning force. Despite these changes, the vortices persisted in the HEAC-0.5% sample at higher temperatures than those in the HEA sample.

## 4. Conclusion

In summary, we fabricated HEA films with compositions of $Ta_{1/6}Nb_{2/6}Hf_{1/6}Zr_{1/6}Ti_{1/6}$ on $Al_2O_3$ substrates, both with and without 0.5% weight carbon additives. The C additives were interstitially introduced into the HEA lattices, which was consistent with the stabilization of the crystal structure and the constancy of the electron-phonon coupling constants. C addition increases the critical transition temperature ($T_c$) and upper critical field ($\mu_0 H_{c2}$) and enhances reflectance in the low-energy region, leading to higher optical conductivity, consistent with decreased electrical resistivity. Superconducting vortices in the HEAC-0.5% sample were observed for the first time via magnetic force microscopic measurements. The shortened magnetic penetration depth in the HEA film with C additives, compared to that of the pure HEA film, could be related to the improvement in superconductivity in the HEA film with C additives. In the future, studies on varying carbon concentrations will be necessary to understand the impact of C additives on structural and physical properties.

**Acknowledgments**

This study was supported by National Research Foundation of Korea (NRF) grants funded by the Korean government (MIST) (RS-2023-00240326 and RS-2023-00220471). J. Hwang acknowledges the support from the National Research Foundation of Korea (2021R1A2C101109811) and the BrainLink program funded by the Ministry of Science and ICT through the National Research Foundation of Korea (2022H1D3A3A01077468). J. Kim acknowledge the support of the National Research Foundation of Korea (NRF) Grant funded by the Korean government (MSIT) (Grant No. NRF-2019R1A2C2090356). W. S. Choi acknowledges the support from the National Research Foundation of Korea (NRF-2021R1A2C2011340).

**AUTHOR DECLARATIONS**

**Conflict of Interest**

The authors have no conflicts to disclose.

**Author Contributions:**

Tien Le and Yeonkyu Lee equally contributed to this work.

**Tien Le:** Data curation (lead); Formal analysis (equal); Investigation (equal); Methodology (equal); Visualization (equal); Writing – original draft (lead); Writing – review & editing (equal). **Yeonkyu Lee:** Data curation (lead); Formal analysis (equal); Investigation (equal); Methodology (equal); Visualization (equal); Writing – original draft (lead); Writing – review & editing (equal). **Dzung T. Tran:** Data curation (equal); **Woo Seok Choi:** Resources (equal); **Won Nam Kang:** Resources (equal); **Jinyoung Yun:** Data curation (equal); **Jeehoon Kim:** Conceptualization (equal); Formal analysis (equal); Funding acquisition (equal); Investigation (equal); Supervision (equal); Writing – original draft (equal); Writing – review & editing (equal); **Jaegu Song:** Data curation (equal); **Yoonseok Han:** Data curation (equal); **Tuson Park:** Funding acquisition (equal); Resources (equal); **Duc H. Tran:** Conceptualization (equal); Data curation (equal); **Soon-Gil Jung:** Conceptualization (equal); Formal analysis (equal); Funding acquisition (equal); Investigation (equal); Supervision (equal); Writing – original draft (equal); Writing – review & editing (equal); **Jungseek Hwang:** Formal analysis (equal); Funding acquisition (equal); Investigation (equal); Supervision (equal); Writing – original draft (equal); Writing – review & editing (lead).

**DATA AVAILABILITY**

The data supporting the findings of this study are available from the corresponding author upon reasonable request.

# References

1. Yeh, J.-W., et al., *Nanostructured High-Entropy Alloys with Multiple Principal Elements: Novel Alloy Design Concepts and Outcomes.* Advanced Engineering Materials, 2004. **6**(5): p. 299-303.
2. Ye, Y.F., et al., *High-entropy alloy: challenges and prospects.* Materials Today, 2016. **19**(6): p. 349-362.
3. Cantor, B., et al., *Microstructural development in equiatomic multicomponent alloys.* Materials Science and Engineering: A, 2004. **375-377**: p. 213-218.
4. Feltrin, A.C., et al., *Review of Novel High-Entropy Protective Materials: Wear, Irradiation, and Erosion Resistance Properties.* Entropy, 2023. **25**(1): p. 73.
5. Kitagawa, J., S. Hamamoto, and N. Ishizu, *Cutting Edge of High-Entropy Alloy Superconductors from the Perspective of Materials Research.* Metals, 2020. **10**(8): p. 1078.
6. Wang, X., W. Guo, and Y. Fu, *High-entropy alloys: emerging materials for advanced functional applications.* Journal of Materials Chemistry A, 2021. **9**(2): p. 663-701.
7. Chen, S.Y., et al., *Phase transformations of HfNbTaTiZr high-entropy alloy at intermediate temperatures.* Scripta Materialia, 2019. **158**: p. 50-56.
8. Yao, Y., et al., *High-entropy nanoparticles: Synthesis-structure-property relationships and data-driven discovery.* Science, 2022. **376**(6589): p. eabn3103.
9. Pickering, E.J. and N.G. Jones, *High-entropy alloys: a critical assessment of their founding principles and future prospects.* International Materials Reviews, 2016. **61**(3): p. 183-202.
10. Jung, S.-G., et al., *High critical current density and high-tolerance superconductivity in high-entropy alloy thin films.* Nature Communications, 2022. **13**(1): p. 3373.
11. Nagase, T., et al., *In-situ TEM observation of structural changes in nano-crystalline CoCrCuFeNi multicomponent high-entropy alloy (HEA) under fast electron irradiation by high voltage electron microscopy (HVEM).* Intermetallics, 2015. **59**: p. 32-42.
12. Koželj, P., et al., *Discovery of a Superconducting High-Entropy Alloy.* Physical Review Letters, 2014. **113**(10): p. 107001.
13. Guo, J., et al., *Robust zero resistance in a superconducting high-entropy alloy at pressures up to 190 GPa.* Proceedings of the National Academy of Sciences, 2017. **114**(50): p. 13144-13147.
14. Zhang, X., et al., *Suppression of the transition to superconductivity in crystal/glass high-entropy alloy nanocomposites.* Communications Physics, 2022. **5**(1): p. 282.
15. Zhang, X., P. Eklund, and R. Shu, *Superconductivity in $(TaNb)_{1-x}(ZrHfTi)_xMo_y$ high-entropy alloy films.* Applied Physics Letters, 2023. **123**(5).
16. von Rohr, F.O. and R.J. Cava, *Isoelectronic substitutions and aluminium alloying in the Ta-Nb-Hf-Zr-Ti high-entropy alloy superconductor.* Physical Review Materials, 2018. **2**(3): p. 034801.
17. Sun, L. and R.J. Cava, *High-entropy alloy superconductors: Status, opportunities, and challenges.* Physical Review Materials, 2019. **3**(9): p. 090301.

18. Chung, K.S., J.H. Luan, and C.H. Shek, *Strengthening and deformation mechanism of interstitially N and C doped FeCrCoNi high entropy alloy.* Journal of Alloys and Compounds, 2022. **904**: p. 164118.
19. He, M.Y., et al., *C and N doping in high-entropy alloys: A pathway to achieve desired strength-ductility synergy.* Applied Materials Today, 2021. **25**: p. 101162.
20. Li, R., et al., *Synthesis, mechanical properties and thermal conductivity of high-entropy (TiTaNbZrMox)(CN) ceramics.* Journal of the European Ceramic Society, 2023. **43**(16): p. 7273-7281.
21. Tanner, D.B., *Optical Effects in Solids*. 2019, Cambridge: Cambridge University Press.
22. Tran, D.T., et al., *Roles of Fe-ion irradiation on MgB2 thin films: Structural, superconducting, and optical properties.* Journal of Alloys and Compounds, 2023. **968**: p. 172144.
23. Kim, G., et al., *Construction of a vector-field cryogenic magnetic force microscope.* Review of Scientific Instruments, 2022. **93**(6).
24. Kim, J., et al., *Direct measurement of the magnetic penetration depth by magnetic force microscopy.* Superconductor Science and Technology, 2012. **25**(11): p. 112001.
25. Cui, Y., et al., *Interstitially carbon-alloyed refractory high-entropy alloys with a body-centered cubic structure.* Science China Materials, 2022. **65**(2): p. 494-500.
26. Zhu, Q., et al., *Structural transformation of MoReRu medium-entropy alloy by carbon addition.* Scripta Materialia, 2022. **210**: p. 114464.
27. Zhu, Q., et al., *Superconducting interstitial MoReRuCx medium-entropy alloys with a hexagonal structure.* Journal of Alloys and Compounds, 2022. **892**: p. 162131.
28. Fereiduni, E. and S.S. Ghasemi Banadkouki, *Ferrite hardening response in a low alloy ferrite–martensite dual phase steel.* Journal of Alloys and Compounds, 2014. **589**: p. 288-294.
29. Ghasemi Banadkouki, S.S. and E. Fereiduni, *Effect of prior austenite carbon partitioning on martensite hardening variation in a low alloy ferrite–martensite dual phase steel.* Materials Science and Engineering: A, 2014. **619**: p. 129-136.
30. Kim, G., et al., *Strongly correlated and strongly coupled s-wave superconductivity of the high entropy alloy Ta1/6Nb2/6Hf1/6Zr1/6Ti1/6 compound.* Acta Materialia, 2020. **186**: p. 250-256.
31. Hidayati, R., et al., *Enhancement of critical current density and critical magnetic field of superconducting medium entropy alloy Nb2/5Hf1/5Zr1/5Ti1/5.* Acta Materialia, 2023. **261**: p. 119420.
32. Kim, J., et al., *Thermal-driven gigantic enhancement in critical current density of high-entropy alloy superconductors.* Journal of Materials Science & Technology, 2024. **189**: p. 60-67.
33. Werthamer, N.R., E. Helfand, and P.C. Hohenberg, *Temperature and Purity Dependence of the Superconducting Critical Field, Hc2. III. Electron Spin and Spin-Orbit Effects.* Physical Review, 1966. **147**(1): p. 295-302.
34. von Rohr, F., et al., *Effect of electron count and chemical complexity in the Ta-Nb-Hf-Zr-Ti high-entropy alloy superconductor.* Proceedings of the National Academy of Sciences, 2016. **113**(46): p. E7144-E7150.

35. Tinkham, M., *Introduction to superconductivity*. 2004, New York: Dover Publication.
36. Ziman, J.M., *Electrons and phonons: the theory of transport phenomena in solids*. 2001: Oxford university press.
37. Bass, J., W.P. Pratt, and P.A. Schroeder, *The temperature-dependent electrical resistivities of the alkali metals.* Reviews of Modern Physics, 1990. **62**(3): p. 645-744.
38. Uporov, S.A., et al., *Pressure effects on electronic structure and electrical conductivity of TiZrHfNb high-entropy alloy.* Intermetallics, 2022. **140**: p. 107394.
39. McMillan, W.L., *Transition Temperature of Strong-Coupled Superconductors.* Physical Review, 1968. **167**(2): p. 331-344.
40. Marik, S., et al., *Superconductivity in a new hexagonal high-entropy alloy.* Physical Review Materials, 2019. **3**(6): p. 060602.
41. Poole Jr, C.P., et al., *Superconductivity Academic Press.* New York, 1995: p. 154.
42. Nazaretski, E., et al., *Direct measurements of the penetration depth in a superconducting film using magnetic force microscopy.* Applied Physics Letters, 2009. **95**(26).
43. Kim, J., et al., *Magnetic penetration-depth measurements of a suppressed superfluid density of superconducting $Ca_{0.5}Na_{0.5}Fe_2As_2$ single crystals by proton irradiation.* Physical Review B, 2012. **86**(14): p. 144509.
44. Le, T., et al., *Influence of aluminum diffusion on $MgB_2$ films grown by hybrid physical–chemical vapor deposition using amorphous aluminum buffers.* Results in Physics, 2024. **58**: p. 107447.
45. Hong, V.T.A., et al., *Probing superconducting gap of the high-entropy alloy $Ta_{1/6}Nb_{2/6}Hf_{1/6}Zr_{1/6}Ti_{1/6}$ via Andreev reflection spectroscopy.* Physical Review B, 2022. **106**(2): p. 024504.
46. Wulferding, D., et al., *Local characterization of a heavy-fermion superconductor via sub-Kelvin magnetic force microscopy.* Applied Physics Letters, 2020. **117**(25).
47. Thompson, J.R., et al., *Vortex pinning and slow creep in high-Jc MgB2 thin films: a magnetic and transport study.* Superconductor Science and Technology, 2005. **18**(7): p. 970.

**Figure Captions and Table**

**Fig. 1.** (a) Schematic of body-centered cubic crystal structure of $Ta_{1/6}Nb_{2/6}Hf_{1/6}Zr_{1/6}Ti_{1/6}$ HEA with C additives expected to act as interstitial defects. (b) XRD patterns of pure HEA and HEAC-0.5% films. Inset shows a magnified view of the (110) peak.

**Fig. 2.** (a)-(b) Scanning electron microscopic (SEM) surface morphologies of HEA and HEAC-0.5% at 3000x magnification, respectively. The film surfaces are smooth with the appearance of droplets. (c) The cross-sectional SEM image shows that the film thickness is approximately 345 nm. (d)-(e) EDS mappings of the HEA and HEAC-0.5% samples show a homogeneous distribution of all elements.

**Fig. 3.** (a) Temperature dependence of electrical resistivity ($\rho$) in magnetic fields perpendicular to surfaces of HEA and HEAC-0.5% films. Determination of superconducting critical temperatures ($T_{cR}$) was based on 50% resistivity transition criterion. (b) Upper critical fields ($\mu_0 H_{c2}$) of HEA and HEAC-0.5% superconducting films. Values are determined using values of magnetic field at $T_{cR}$. Black dashed and red dot-dot-dashed lines indicate linear fits, which are used to estimate $d(\mu_0 H_{c2})/dT$ at $T_{cR}$.

**Fig. 4.** (a) Temperature dependence of electrical resistivity ($\rho$) of (a) HEA and (b) HEAC-0.5% films. Solid red lines are fits using Bloch–Grüneisen expression. Fitting parameters are Debye temperature ($\Theta_{Debye}$) and residual resistivity ($\rho_0$).

**Fig. 5.** (a) Measured reflectance spectra at 300 K in a wide spectral range from 100 to 40,000 cm$^{-1}$, (b) and (c) Real parts of the optical conductivity for HEA and HEAC-0.5%, respectively, obtained from Kramers-Kronig analysis.

**Fig. 6.** (a) 20 × 20 μm$^2$ MFM image of Nb reference film obtained at $T$ = 4.25 K with tip-sample distance of 400 nm. (b) and (c) 12 × 12 μm$^2$ MFM images of HEA and HEAC-0.5%

films taken at $T = 4.25$ K with tip-sample distance of 300 nm. (d) Line profiles across vortices as indicated by white lines in (a) and (c). (e) and (f) Meissner curves of HEA and HEAC-0.5% films matched with that of Nb film. Measured positions are marked by red dots in (a), (b), and (c).

**Fig. 7**. MFM images of HEA((a)-(c)) and HEAC-0.5%((d)-(f)) films obtained with increasing temperature from 4.25 K to 6 K. All MFM images are taken at tip-sample distance of 400 nm with scan size of $12 \times 12$ μm$^2$.

**Table 1.** Statistics table for EDS results of the elemental distribution in the HEA and HEAC-0.5%.

**Table 2.** Fitting parameters for real part of the optical conductivity ($\sigma_1(\omega)$).

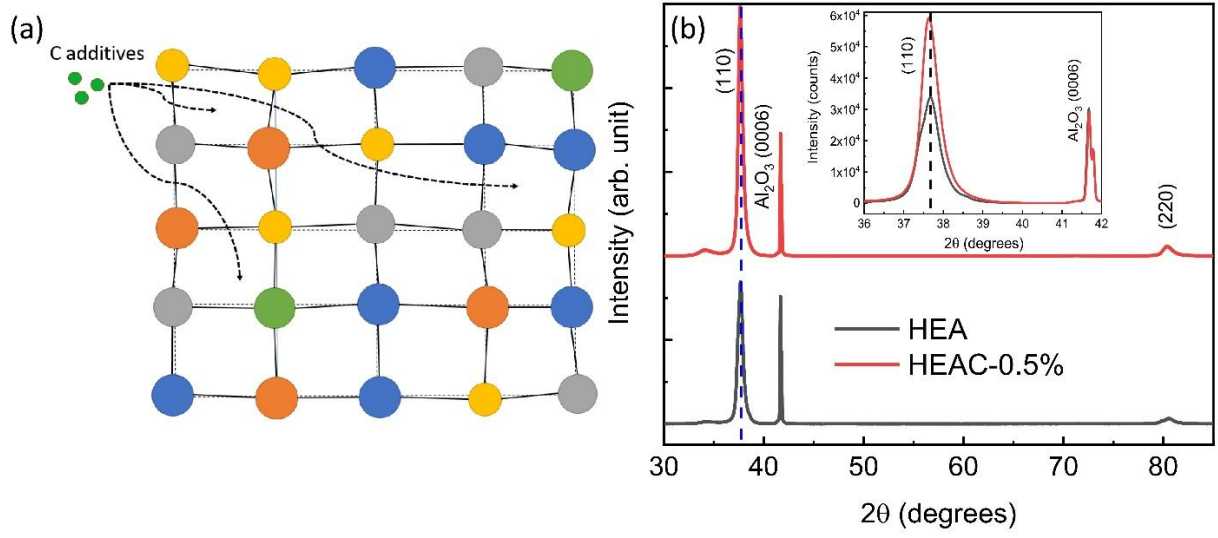

**Fig. 1**

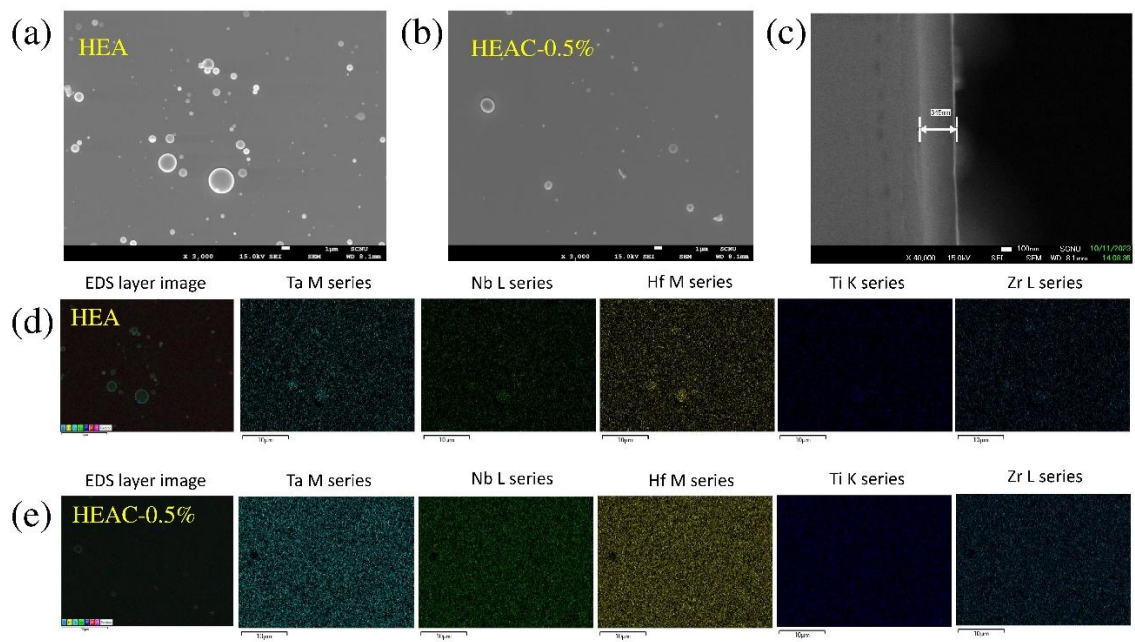

**Fig. 2**

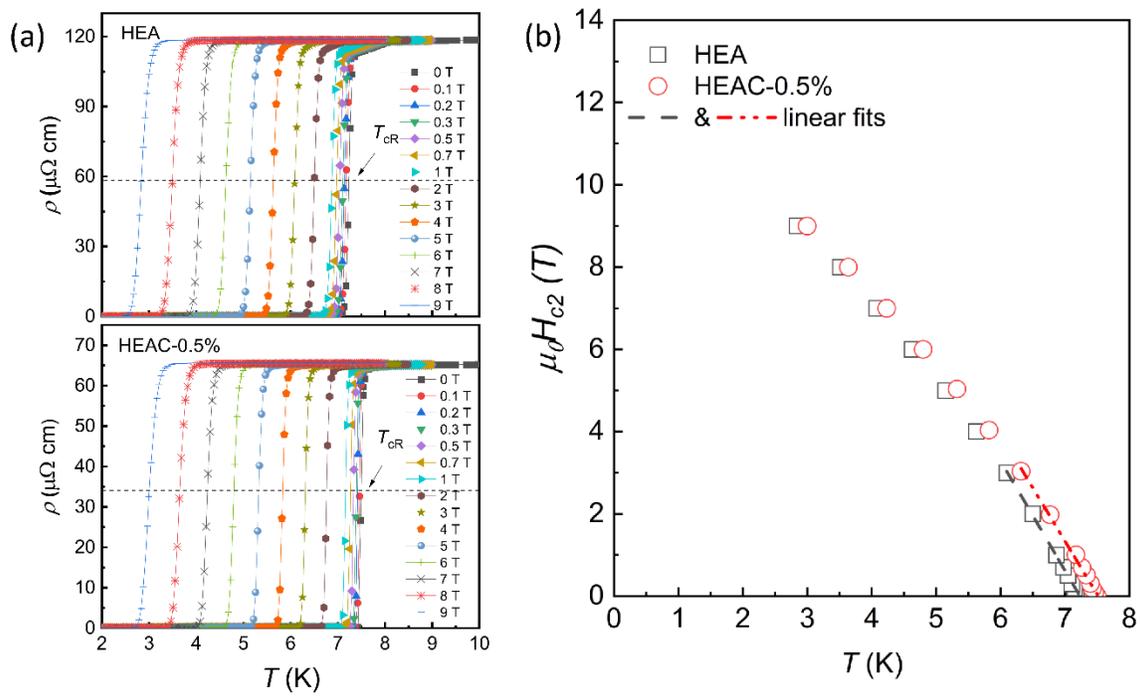

**Fig. 3**

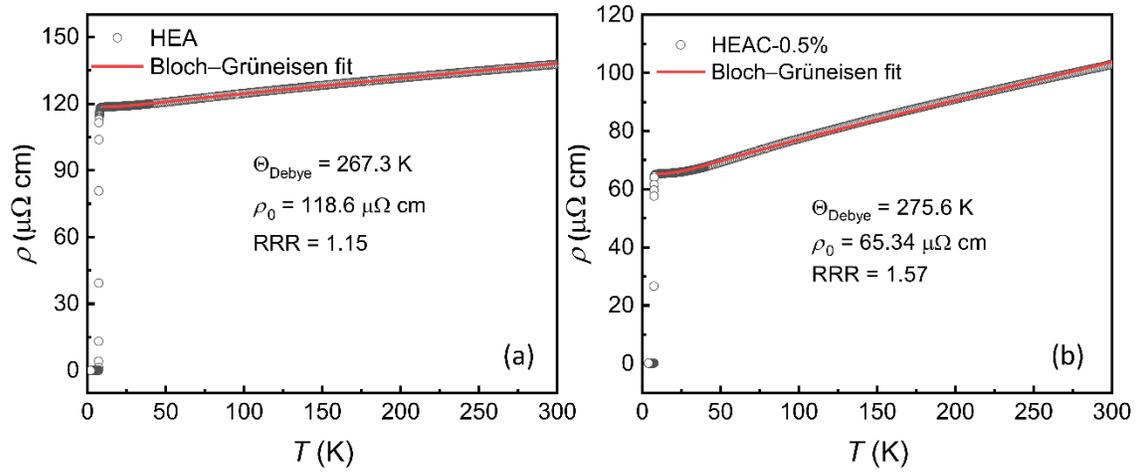

**Fig. 4**

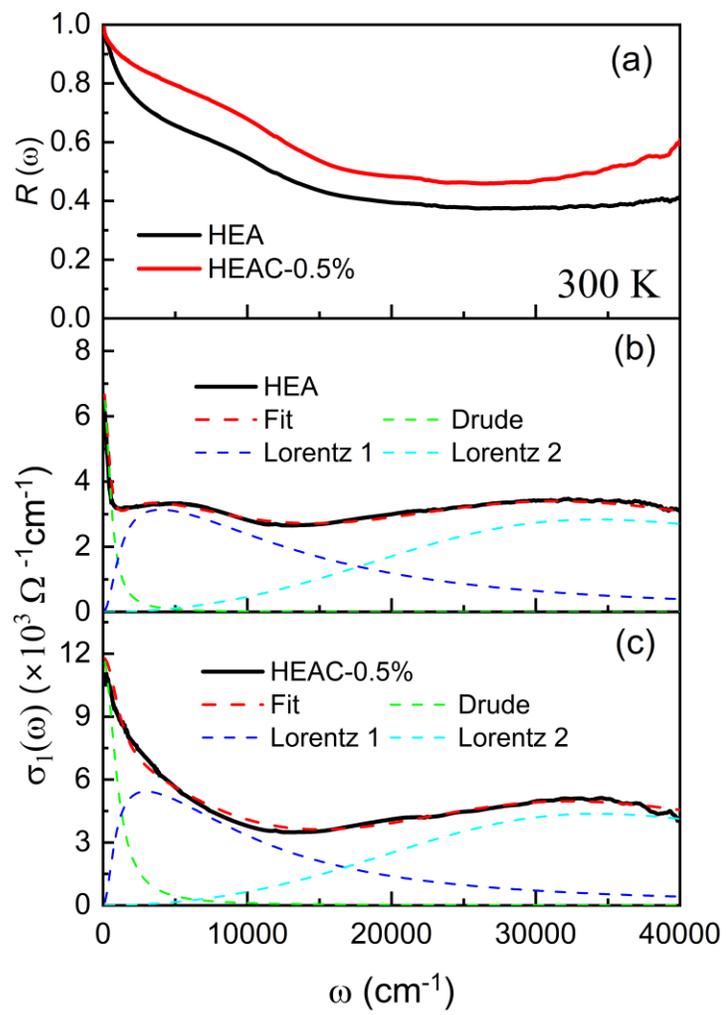

**Fig. 5**

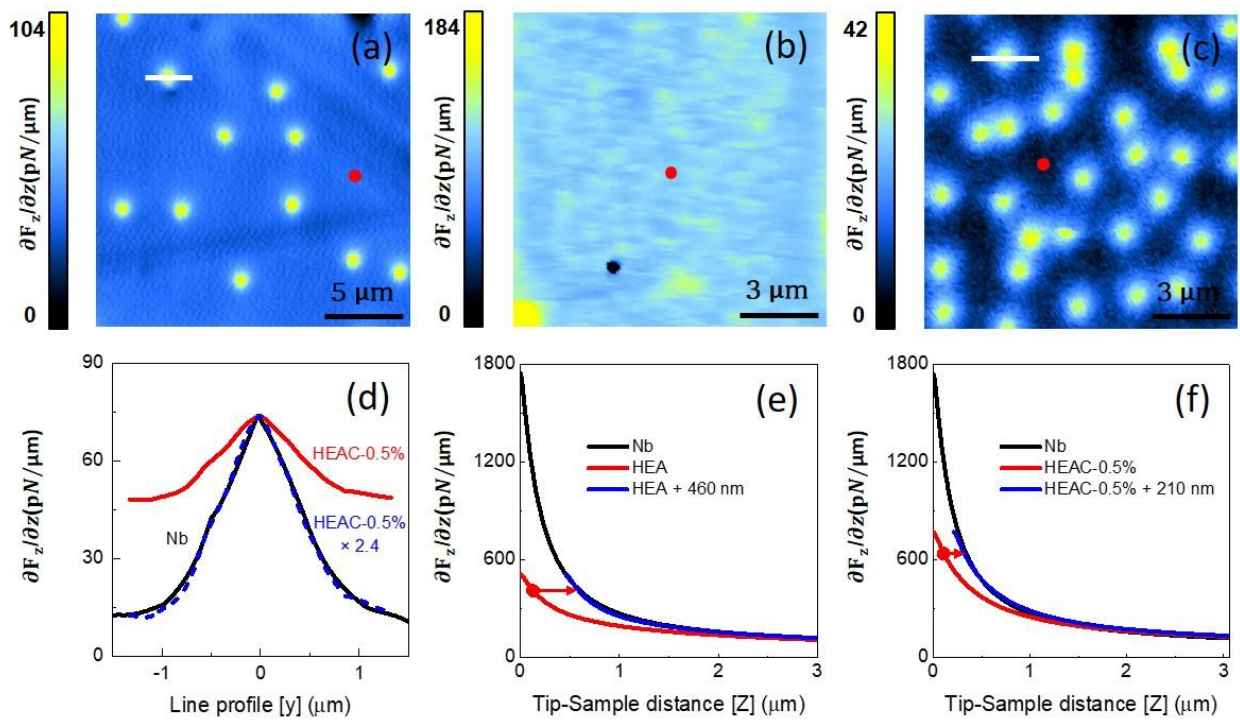

**Fig. 6**

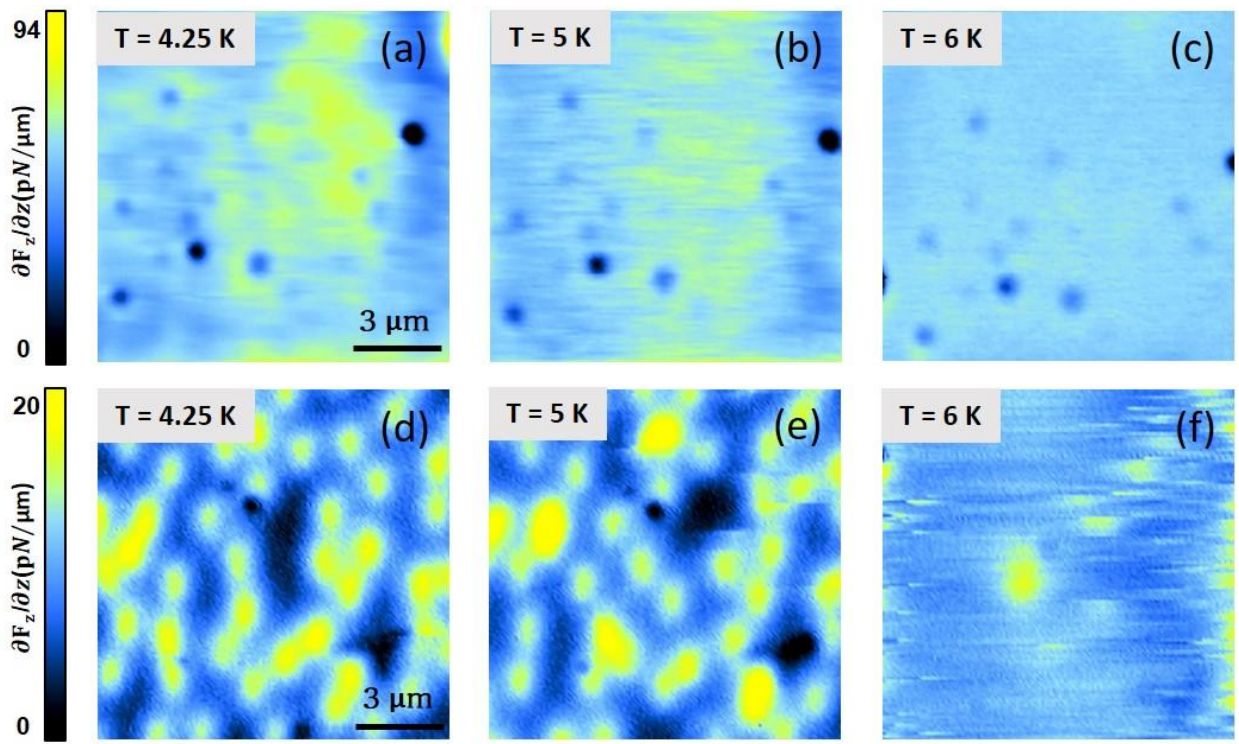

**Fig. 7**

| Element | Atomic % ||
|---|---|---|
| | HEA | HEAC-0.5% |
| Ta | 11.75 | 13.95 |
| Nb | 37.66 | 34.25 |
| Hf | 18.49 | 21.52 |
| Ti | 17.41 | 14.96 |
| Zr | 14.69 | 15.32 |

**Table 1.**

| Sample | $\Omega_{p,D}$ (cm$^{-1}$) | $1/\tau_{imp,D}$ (cm$^{-1}$) | $\Omega_{p,1}$ (cm$^{-1}$) | $\omega_1$ (cm$^{-1}$) | $\gamma_1$ (cm$^{-1}$) | $\Omega_{p,2}$ (cm$^{-1}$) | $\omega_2$ (cm$^{-1}$) | $\gamma_2$ (cm$^{-1}$) |
|---|---|---|---|---|---|---|---|---|
| HEA | 15105 | 556 | 53163 | 4002 | 15030 | 89779 | 34190 | 47373 |
| HEAC-0.5% | 26487 | 989 | 61187 | 2974 | 11501 | 105000 | 34175 | 44536 |

**Table 2.**